\documentstyle[aps,prl,twocolumn,graphicx]{revtex}

\begin{document}
\draft

\title{Observation of three-magnon light scattering in CuGeO$_3$}

\author{G. Els, P.H.M. van Loosdrecht, P. Lemmens, H. Vonberg,
G. G\"untherodt}
\address{
II. Physikalisches Institut, RWTH-Aachen, Templergraben 55, 52056 Aachen,
Germany.}
\author{G. S. Uhrig}
\address{Institut f\"ur Theoretische Physik, Universit\"at zu K\"oln,
 Z\"ulpicherstra\ss e 77, 50937 K\"oln, Germany.}
\author{O. Fujita, J. Akimitsu}
\address{Department of Physics, Aoyama Gakuin University, 6-16-1 Chitsedai,
Setagaya-ku, Tokyo 157, Japan.}
\author{G. Dhalenne, A. Revcolevschi}
\address{Laboratoire de Chimie des Solides, Universit\'{e} de Paris-Sud,
b\^atiment 414, F-91405 Orsay, France}
\date{submitted May 5, 1997, Phys. Rev. Lett. in press.}
\maketitle

\begin{abstract}
Temperature and magnetic field dependent polarized Brillouin spectra
of CuGeO$_3$\ are reported. In addition to a bound singlet state at 
30~cm$^{-1}$, a new feature has been observed at 18~cm$^{-1}$.
This feature is interpreted in terms of a novel three-magnon light 
scattering process between excited triplet states.
\end{abstract}
\pacs{75.40.Gb, 78.35.+e, 75.50.Ee, and 75.10.Jm}

\narrowtext
The unusual properties of low dimensional antiferromagnetic Heisenberg
spin-chains,
such as the occurence of the Haldane gap \cite{halda83},
low energy quantum fluctuations and dominant spin-wave continua
\cite{mulle81},
and the magneto-elastic spin-Peierls (SP) transition \cite{pytte74b},
attracted the interest of both experimental and theoretical physicists
over the past two decades. Recently, this interest
has been boosted once more by the discovery \cite{hase93a} of a SP transition
in the inorganic compound CuGeO$_3$. This inorganic nature of CuGeO$_3$
tremendously increased the
possibilities for experimental studies of a SP system\cite{bouch96}.
These in return intensified the theoretical efforts to understand the
properties of CuGeO$_3$\ and of frustrated dimerized $S=1/2$\ antiferromagnetic
spin-chains in general.

To first approximation, CuGeO$_3$\ is  described as a
frustrated quasi one-dimensional $S=1/2$\ isotropic Heisenberg
antiferromagnet coupled to the elastic degrees of freedom of the lattice.
Due to this coupling the system undergoes a SP transition
at $T_{\rm\scriptstyle SP}=14$\ K to a dimerized non-magnetic ground state.
Frustration in this system
arises from competing antiferromagnetic nearest neighbor (nn) and
next-nearest neighbor (nnn) interactions ($J_{\rm nn}=120-160$\ K,
$\alpha=J_{\rm nnn}/J_{\rm nn}=0.24-0.37$ \cite{riera95,casti95,buchn96}).
The influence of  interchain coupling \cite{uhrig97a} will be discussed
later in this Letter.

The spin-Peierls transition results from the instability of spin chain
systems towards dimerization.
The gain in magnetic energy overcompensates the
necessary elastic energy $\propto \delta^2$ ($\delta$:
dimerization amplitude).
For $\alpha < \alpha_{\rm\scriptstyle c}$
($\alpha_{\rm\scriptstyle c} = 0.241$
\cite{julli83}) this gain is proportional to $\delta^{4/3}$
\cite{cross79}.
For  $\alpha > \alpha_{\rm\scriptstyle c}$ the gain is even linear in
$\delta$
since the magnetic system in itself breaks already the translational
symmetry \cite{halda82b,julli83}. In this case the magnetic system
displays also a
relatively small energy gap $\Delta_{\rm\scriptstyle frus}$
{\em without} lattice dimerization. 

Inelastic light scattering (ILS) has proven to be a sensitive technique to
study magnetic excitations in CuGeO$_3$\ \cite{kuroe94a,loosd96b,lemme96,loa96}.
It is the aim of the present Letter to show the existence of
a so far unobserved type of magnetic scattering process in gapful
low-dimensional spin systems. This scattering process was seen 
by temperature and magnetic field dependent polarized Brillouin
scattering experiments in the SP phase of CuGeO$_3$.

The single crystals used in this study were grown by two different groups
\cite{revco93,GROWTHJAPAN} using a travelling floating zone method.
We did not note any qualitative differences between crystals from the two 
sources. The crystals were cleaved along the (100) planes to obtain a virgin 
surface and mounted in a He flow cryostat (stabilized within $\pm$0.1~K).
Polarized Brillouin spectra were recorded in a 90$^{\rm o}$\ geometry using 
a Sandercock type tandem Fabry-Perot spectrometer, with the 514.5~nm line of 
an Ar$^+$-laser as excitation source (incident fluence $\leq$140~W/cm$^2$).
Because of the transparency of the CuGeO$_3$ samples heating effects were estimated
to be less than 0.5~K. The advantage of Brillouin spectroscopy is its high contrast 
close to the laser line. For the experiment presented a particular small mirror 
spacing of 100~$\mu$m is used to achieve the necessary large free spectral range 
of 50~cm$^{-1}$. To ease comparison with Raman spectra we choose cm$^{-1}$ as 
energy scale (1~cm$^{-1}=30$~GHz$~=0.125$~meV).

\begin{figure}
\centerline{\includegraphics[width=7.5cm]{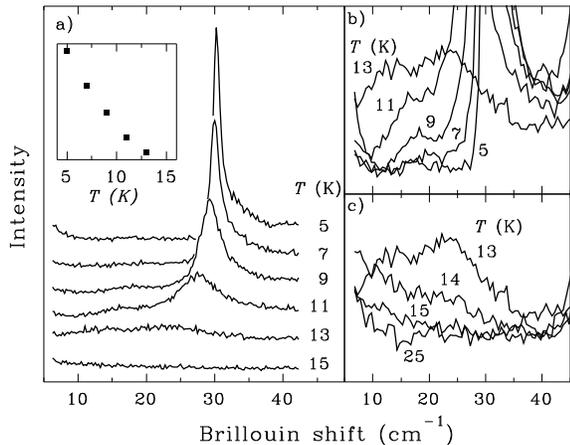}}
\caption{\label{boundstate}
(ZZ) polarized Brillouin spectra for several temperatures 
showing the singlet bound state response (a) (the curves have been given 
an offset for clarity); a thermally induced low energy 
scattering process below $T_{\rm\scriptstyle SP}$\ (b);
and the disappearance of the scattering intensity for 
$T>T_{\rm\scriptstyle SP}$\ (c).
The inset in (a) shows the $T$-dependence of the peak 
intensity of the singlet bound state response.}
\end{figure}

In ILS the opening of the magnetic gap in the SP phase is evidenced by the 
appearance of a sharp asymmetric peak in the spectrum around 30~cm$^{-1}$\ 
\cite{kuroe94a,loosd96b}. The appearance of this mode is 
depicted in Fig.\ \ref{boundstate}a, which displays (ZZ) polarized Brillouin 
spectra (Z$\parallel$chains) for several temperatures in the vicinity of the 
SP transition. At the lowest temperature a clear, asymmetric, resolution 
limited peak is observed at 30~cm$^{-1}$. With increasing temperature the 
observed peak shifts to lower energies 
\noindent
and strongly broadens. This is 
evidently associated with the closure of the SP gap upon approaching the 
SP transition at 14~K. The inset of Fig.\ \ref{boundstate}a shows the peak 
intensity of the singlet bound state response as a function of temperature. 
An extrapolation of the peak intensity beyond 13~K is in good agreement with 
the SP transition temperature ($T_{\rm\scriptstyle SP}=14$~K). 

\begin{figure}
\centerline{\includegraphics[width=7.5cm]{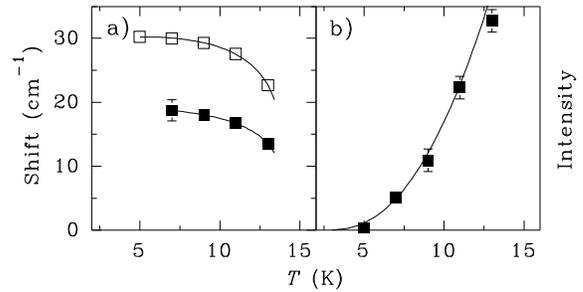}}
\caption{\label{tempdep}a) Temperature dependence of the energy of the
singlet bound state (open symbols) and the 18~cm$^{-1}$ (filled symbols) 
response. Solid lines are guides to the eye. b) Temperature dependence of 
the intensity of the 18~cm$^{-1}$\ shoulder (filled symbols). The solid 
line demonstrates the activated $T$-dependence of the intensity of the 
18~cm$^{-1}$\ response (see text).}
\end{figure}

It is evident from the Brillouin spectra in Fig.\ \ref{boundstate} that the 30~
cm$^{-1}$\ mode is not the only low energy excitation. This is more clearly 
depicted in Figs.\ \ref{boundstate}b and c, which show the  data on an expanded 
intensity scale. One observes a low energy shoulder in spectra between 
$T_{\rm\scriptstyle SP}$\ and 7~K. The $T$-dependence of the frequency 
and peak intensity of this shoulder are shown in Figs.\ \ref{tempdep}a and b 
(filled symbols), respectively. At $T=13$~K, the shoulder is found at about
 13~cm$^{-1}$. As $T$ is decreased, it is shifted towards higher energies, but 
simultaneously looses its intensity until it finally becomes unobservable at 
about 5~K. Apparently, the $T$-dependence of the frequency of this shoulder is 
similar as that found for the 30~cm$^{-1}$\ singlet bound state response 
(Fig.\ \ref{tempdep}a, open symbols). Furthermore, we did not find any intensity 
on the anti-Stokes side of the spectra consistent with the thermal suppression 
of the intensity 
$I_{\rm anti-Stokes}/I_{\rm Stokes}\propto \exp(-\hbar\omega/(k_{\rm B}T))$
in this frequency and temperature range.

The unusual $T$-dependence of the intensity of the 18~cm$^{-1}$\ shoulder 
strongly indicates that the scattering process 
involved is due to transitions between {\em excited} states. It implies also 
that this shoulder is distinct from the structure observed in Raman scattering 
experiments in the same frequency region, but at a rather low temperature 
$T=2$~K, which has tentatively been assigned to one-magnon 
scattering\cite{loa96}. We did not find any signature of the latter mode.
 The shoulder observed here is fully (ZZ) polarized.
No scattering intensity could be 
\noindent
observed in crossed (ZY) or
circular (Z\,c) polarization of the scattered light. Also
no magnetic field dependence could be found up to  $6$~T.
These facts rule out a one-magnon process which should have
an anti-symmetric Raman  tensor and should display Zeeman splitting.

\begin{figure}
\centerline{\includegraphics[width=7.5cm]{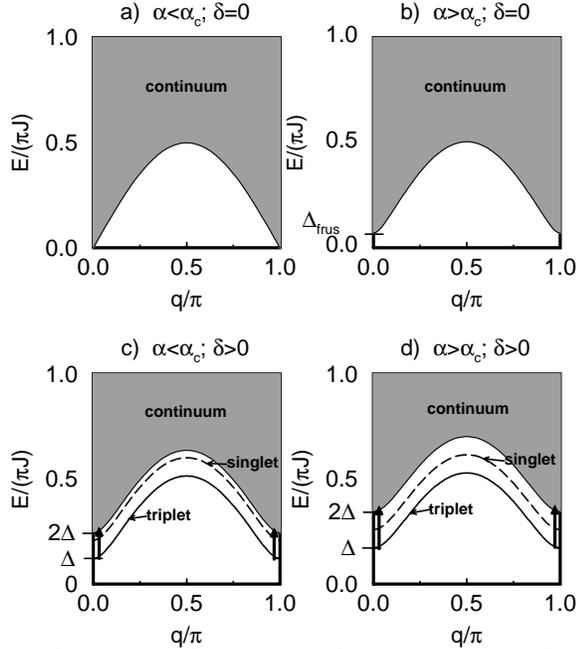}}
\caption{\label{schemaneu}
Schematic representations of the magnetic excitation
spectrum expected for a $S=1/2$\ Heisenberg antiferromagnetic
chain; with dimerization (c) and (d); with frustration (b) and (d).
Thin solid line: onset of the continuum of $S=1$ or $S=0$ excitations.
Shaded area: continuum (No statement on spectral weights is made).
Thick solid line: elementary magnon excitation $S=1$.
Dashed line: bound state singlet excitations. 
The vertical arrows describe the three-magnon scattering (see text).}
\end{figure}

For a better understanding let us recall the low lying excitations in 
dimerized frustrated spin chains (see Fig.\ \ref{schemaneu}). For 
$T>T_{\rm\scriptstyle SP}$\ we have no dimerization ($\delta=0$) and, 
depending on the value of $\alpha$, Fig.\ \ref{schemaneu}a \cite{mulle81} 
or Fig.\ \ref{schemaneu}b \cite{shast81} applies. In the dimerized SP 
phase ($T<T_{\rm\scriptstyle SP}$), we have $\delta>0$ and a gap $\Delta$ 
opens between the singlet ground state and the lowest elementary triplet 
excitations ($S=1$, ``magnons''). These triplets (Fig.\ \ref{schemaneu}, 
thick solid line) are again separated by a gap $\Delta$ from the two-magnon 
continuum (Fig.\ \ref{schemaneu}, shaded area) starting at 2$\Delta$
\cite{uhrig96b}. For $\delta$ not too small, there is no qualitative 
difference between $\alpha<\alpha_{\rm\scriptstyle c} $ and 
$\alpha>\alpha_{\rm\scriptstyle c} $, see Figs.\ \ref{schemaneu}c, d. In 
addition to the $S=1$ excitations a well-defined singlet mode is predicted 
below the continuum (Figs.\ \ref{schemaneu}c and d, dashed lines) 
\cite{uhrig96b,bouze97a}. It can be viewed  as a singlet bound state of two 
antiparallel magnons \cite{uhrig96b}. The energy difference from the continuum 
onset at $2\Delta$ is its binding energy. Larger dimerization lowers, whereas 
larger frustration enhances the singlet binding effect \cite{uhrig96b,loosd97a}.

The existence of a continuum above $T_{\rm\scriptstyle SP}$\ in CuGeO$_3$\ 
is confirmed by Raman \cite{loosd96b} and inelastic neutron \cite{arai96} 
scattering experiments. The opening of a gap in the magnetic excitation 
spectrum was experimentally verified in CuGeO$_3$\ using a 
variety of methods \cite{hase93a,nishi94,brill94,loosd96a}.
First neutron experiments on the magnetic excitation spectrum of CuGeO$_3$\ 
in the SP phase showed the existence of a single branch of well defined 
elementary triplet excitations \cite{nishi94}. Recently, inelastic neutron 
scattering experiments revealed that these elementary triplet excitations 
are separated from a continuum starting at about twice the gap \cite{ain97}. 
The singlet bound state below the onset of the continuum is visible in ILS data 
at 30~cm$^{-1}$ (see Fig.\ \ref{boundstate}a) relatively close to the 
continuum onset \cite{kuroe94a,loosd96b}. Initially it was suggested that 
this mode is due to a bound two-magnon state \cite{kuroe94a}.
Other authors \cite{loosd96b,lemme96,loa96} identified this 
\noindent
peak as due to a 
direct two-magnon scattering process from the SP gap excitations, 
based on the near coincidence of this peak with twice the SP gap.
Later, however, Raman experiments under pressure confirmed the former 
hypothesis of a bound two-magnon singlet state, as its binding 
energy strongly increases upon increasing pressure due to an 
enhancement of the frustration $\alpha$\cite{loosd97a}.

We now propose an explanation for the 18~cm$^{-1}$\ shoulder which is 
consistent with the above observations. Within standard Fleury-Loudon 
theory \cite{fleur68} for ILS in magnetic solids the transition operator 
for a dimerized and frustrated spin chain (see also \cite{muthu96}) is 
of the type 
\begin{equation}
\label{een}
R= \sum_{i} (1+(-1)^i\delta) {\rm\bf S}_i\cdot{\rm\bf S}_{i+1}
 +  \gamma {\rm\bf S}_i\cdot{\rm\bf S}_{i+2}  \ .
\end{equation}
This operator $R$ conserves the total spin since it is a scalar.
In order to gain insight into possible scattering processes it is allowed to
work in the limit of strong dimerization with $\delta$ close to unity.
This point of view has proven very useful for a qualitative understanding
\cite{uhrig97a,uhrig96b}. In this limit the ground state is a product of 
local singlets at the strong bonds (henceforth called dimers).
A magnon excitation consists of one triplet at one of the dimers.
Due to the residual weak bonds these triplets acquire a dispersion. 
The application of $R$ in (\ref{een}) on the ground state creates two 
triplets on adjacent dimers combined to $S=0$. Thus a light absorption 
experiment at $T=0$ probes essentially the two-magnon subspace displaying 
a continuum and a bound singlet state below the continuum onset \cite{uhrig96b}.

At finite temperature we may also start from the situation where one magnon 
is already excited. The application of $R$ at $T\neq0$ might 
(i) create two {\em additional} adjacent triplets somewhere (two-magnon scattering) or 
(ii) convert one triplet into two adjacent triplets.
The latter process is what we call a three-magnon scattering process.
Spin conservation requires that these two triplets are combined to $S=1$.

For illustration let us consider four spins ${\rm\bf 
S}_1$, ${\rm\bf S}_2$, ${\rm\bf S}_3$, and ${\rm\bf S}_4$ 
forming two singlets $s_{12}$ and $s_{34}$. We denote the singlet product
as $|s,s\rangle$ omitting the site indices.
This state is the ground state for $\delta\to 1$. We 
use $R = {\rm\bf S}_2\cdot{\rm\bf S}_3$ as Raman operator with 
$R=S^z_2 S^z_3 + (1/2) (S^+_2 S^-_3 + S^-_2 
S^+_3)$. Application of $R$ on $|s,s\rangle$ yields 
\begin{equation}
\label{raman1}
R|s,s\rangle = (1/4) (-|t_0, t_0\rangle + |t_1, t_{-1}\rangle +
 |t_{-1} t_{1}\rangle )
\end{equation}
where $t_{0,\pm1}$ denotes the triplet state with corresponding $S^z$
 component.  The total spin before and after the application of $R$ is 
zero. Process (\ref{raman1}) creates two elementary magnons and corresponds to 
the $T=0$ situation and to process (i) above. If we choose $|t_0,s\rangle$ as 
initial state to describe process (ii), the result is 
\begin{equation}
\label{raman2}
R|t_0,s\rangle = (1/4) (-|s, t_0\rangle + |t_1, t_{-1}\rangle -
 |t_{-1}, t_{1}\rangle )\ .
\end{equation}
Here, the total spin before and after the application of $R$ is unity.
The first term in (\ref{raman2}) does not change the number of 
magnons in the system. The other terms induce a transition from
one to two magnons thus incrementing the number of magnons by one.
This is the three-magnon process (ii). Note that this process is lost in any 
bosonic description (Holstein-Primakov or bond-operator)
of antiferromagnets where only the quadratic terms are kept. For this 
reason we conclude that the three-magnon process is a generic feature
in gapful low-dimensional spin liquids which are {\em not}
described by N\'eel-type states.

Treating the magnons like independent bosons the
$T$-dependence of the intensity of the three-magnon process is given by
\begin{equation}
\label{intensity}
I \propto n(\omega_{\rm\scriptstyle i})
(1+ n(\omega_{\rm\scriptstyle f1}))(1+ n(\omega_{\rm\scriptstyle f2}))
\end{equation}
where $n(\omega)$ is the Bose distribution function and
$\omega_{\rm\scriptstyle i}$,
$\omega_{\rm\scriptstyle f1}$, and  $\omega_{\rm\scriptstyle f2}$
are the magnon
energies of  the initial magnon and the two final magnons, respectively.
For $T\ll \Delta\approx23$\ K one has 
$I \propto n(\omega_{\rm\scriptstyle i}) $, which for $T=0$\ 
vanishes since  there are no  excited magnons in the system anymore.

Returning to the 18~cm$^{-1}$\ response observed in the Brillouin spectra 
(Fig.\ \ref{boundstate}) we interprete this structure now as the {\em onset 
of a continuum} of momentum conserving vertical ($\Delta\vec{q}=0$) 
transitions from the lowest magnon branch (thick line in Fig.\ \ref{schemaneu}) 
to the two-magnon continuum of $S=1$ excitations (shadded area in 
Fig.\ \ref{schemaneu}), see vertical arrows in Figs.\ \ref{schemaneu}c and d. 
The three-magnon scattering relies on the same operator $R$ as does the rest 
of the Brillouin spectrum. It therefore obeys the same selection rules, in 
excellent agreement with experiment and conserves the total spin and each spin 
component. Thus there is no shift or splitting in a magnetic field as observed 
experimentally. In view of the low temperatures of the experiments, one expects 
that the major contribution to the scattered intensity comes from transitions 
with an initial state which has an energy comparable to the SP gap 
$\Delta=2.1$~meV. The onset of the observed scattering is therefore determined 
by the energy difference between the onset of the continuum at about 
$2\Delta=4.2$~meV \cite{ain97} and the SP gap which is again $\Delta=2.1$~meV. 
This is in good agreement with the data in Fig.\ \ref{boundstate}. The 
temperature dependence of the intensity at 18~cm$^{-1}$\ (Fig.\ \ref{tempdep}b) 
is also found to agree well with the expected dependence (\ref{intensity}). 
This is shown in Fig.\ \ref{tempdep}b by the solid line which is calculated using
$\omega_{\rm\scriptstyle i} = \omega_{\rm\scriptstyle f1} =
\omega_{\rm\scriptstyle f2} = \Delta =2.1$~meV.
Since the softening of the 18 cm$^{-1}$\ structure follows the same behavior as both 
the singlet bound state response and the spin-Peierls gap (see Fig.\ \ref{tempdep}a), one may 
conclude that all excitations depicted in Fig.\ \ref{schemaneu} renormalize in the 
same manner until they merge with the continuum of the uniform isotropic $S=1/2$\ 
Heisenberg antiferromagnet \cite{mulle81,loosd96b,arai96} at the phase transition.

So far we discussed the energies of the three-magnon process in CuGeO$_3$\ on the 
basis of a $d=1$ model. In order to demonstrate that the interchain coupling
does not change the picture qualitatively we refer to the $d>1$ dispersion
calculated in \cite{uhrig97a}, see Fig.\ 4 therein. If we start from the magnon 
at the zone center $\vec{q}_{\rm\scriptstyle i}=(0,0,0)$ with 
$\omega_{\rm\scriptstyle i} = 2.5$~meV we may induce a momentum conserving 
transition to  the magnons with minimum energy
$\omega_{\rm\scriptstyle f1} = \omega_{\rm\scriptstyle f2} = \Delta = 2.1$~meV
at $\vec{q}_{\rm\scriptstyle f1}=(0,1,1/2)$
and at $\vec{q}_{\rm\scriptstyle f2}= -\vec{q}_{\rm\scriptstyle f1}$.
This leads to an onset at $2\times 2.1 - 2.5 = 1.7$~meV or 14~cm$^{-1}$.
All transitions at other points in the Brillouin zone lead to
larger values for the onset. Inspection of Fig.\ \ref{boundstate} shows that
the onset at 14~cm$^{-1}$\ is also (perhaps even better) compatible with the
experiment. Thus our explanation is not restricted to the (simplified) $d=1$ 
picture of CuGeO$_3$.

In conclusion, we reported on a low energy transition in CuGeO$_3$\
observed in Brillouin scattering. This transition is
assigned to a novel three-magnon scattering process between
the lowest triplet branch and the continuum of triplet states in CuGeO$_3$\
in excellent agreement with the experimental results for energy,
temperature dependence, magnetic field dependence, and polarization
selection rules. The three-magnon process is expected to be 
a generic feature in gapful, low-dimensional spin systems.

We are grateful for helpful discussions with G. Bouzerar, P. Fumagalli, B.
Hillebrands, J.R. Sandercock, F. Sch\"onfeld, and M. Udagawa. 
This work was supported by the DFG through SFB 341 and by the BMBF Fkz.~13N6586/8.
Laboratoire de Chimie des Solides is a "Unit\'{e} de Recherche Associe\'{e} CNRS 
n$^o$ 446".


\end{document}